\documentclass[12pt]{article}

\usepackage{amsmath,amsfonts,theorem}
\usepackage{bm}
\usepackage{natbib}
\usepackage[dvipdf]{graphicx}
\usepackage{amssymb}
\usepackage{epsfig}
\usepackage{color}
\usepackage{latexsym}
\usepackage{multicol}
\usepackage{setspace}

\usepackage{graphicx}
\usepackage{caption}
\usepackage{subcaption}
\usepackage{authblk}

\def\v{{\varepsilon}}

\renewcommand{\baselinestretch} {1.30}
\newtheorem{theorem}{Theorem}
\newtheorem{lemma}{Lemma}
\newtheorem{corollary}{Corollary}

\theoremheaderfont{\hskip\parindent\normalfont}
\theorembodyfont{\normalfont}
\newtheorem{example}{\itshape Example}

\newcommand{\bw}{\mathbf{w}}

\newcommand{\cA}{\mathcal{A}}
\newcommand{\cB}{\mathcal{B}}
\newcommand{\cE}{\mathcal{E}}

\newcommand{\cS}{\mathcal{S}}

\newcommand{\cI}{\mathcal{I}}

\newcommand{\bmu}{\boldsymbol{\mu}}

\newcommand{\bzero}{\boldsymbol{0}}
\newcommand{\bone}{\boldsymbol{1}}

\begin{document}

\numberwithin{equation}{section}
\renewcommand{\baselinestretch}{1.5}

\title{A super scalable algorithm for short segment detection}

\author[1]{Ning Hao}
\author[1]{Yue Selena Niu}
\author[2]{Feifei Xiao}
\author[3]{Heping Zhang}
\affil[1]{University of Arizona}
\affil[2]{University of South Carolina}
\affil[3]{Yale University}
\date{}                     
\setcounter{Maxaffil}{0}
\renewcommand\Affilfont{\itshape\small}

\maketitle

\begin{abstract}
In many applications such as copy number variant (CNV) detection, the goal is to identify short segments on which the observations have different means or medians from the background. Those segments are usually short and hidden in a long sequence, and hence are very challenging to find. We study a super scalable short segment (4S) detection algorithm in this paper. This nonparametric method clusters the locations where the observations exceed a threshold for segment detection. It is computationally efficient and does not rely on Gaussian noise assumption. Moreover, we develop a framework to assign significance levels for detected segments. We demonstrate the advantages of our proposed method by theoretical, simulation, and real data studies.

\end{abstract}

\section{Introduction}
Chromosome copy number variant (CNV) is a type of structural variation with abnormal copy number changes involving DNA fragments \citep{freeman2006copy,feuk2006structural}. CNVs result in gains or losses of the genome, therefore interfering downstream functions of the DNA contents. Accounting for a substantial amount of genetic variation, CNVs are considered to be a risk factor for human diseases. Over the past decade, advances in genomic technologies have revealed that CNVs underlie many human diseases, including autism \citep{pinto2010functional}, cancer \citep{fanale2013analysis}, schizophrenia \citep{castellani2014copy}, and major depressive disorder \citep{o2014rare}. It is fundamental to develop fast and accurate CNV detection tools.

A variety of statistical tools have been developed to discover structural changes in CNV data during last 20 years. Popular algorithms include circular binary segmentation \citep{olshen2004circular}, the fused LASSO \citep{tibshirani2008spatial}, likelihood ratio selection \citep{jeng2010optimal}, and screening and ranking algorithm \citep{niu2012screening}. Some other change-point detection tools such as wild binary segmentation \citep{fryzlewicz2014wild} and simultaneous multiscale changepoint estimator \citep{frick2014multiscale} can be also applied to CNV data. See \cite{niu2016multiple} for a recent review on modern change-point analysis techniques. A majority of existing methods are based on Gaussian assumption, although quantile normalization \citep{xiao2014modified} or local median transformation \citep{tony2012robust} can be used for normalization. The computational complexity is also of concern for some of the existing methods as the modern technologies produce extraordinarily big data. In spite of some fast algorithms \citep{wang2007penncnv}, few algorithms are known to possess both computational efficiency and solid theoretical foundation. Moreover, with only a few exceptions \citep{hao2013multiple,frick2014multiscale}, the existing methods focus on detection whereas not offering statistical inference. For these reasons, we develop a fast nonparametric method for CNV detection with theoretical foundation and the opportunity of conducting statistical inference.

In this paper, we model the CNVs as short segments with nonzero height parameters, which are sparsely hidden in a long sequence. The goal is to identify those segments with high probability and, moreover, to assess the significance levels for detected segments. In particular, we propose a scalable nonparametric algorithm for short segment detection. It depends on only the ranks of the absolute values of the measurements and hence requires minimal assumptions on the noise distribution. A short segment may be present when there are a large enough number of observations exceeding a certain threshold on a short segment; for instance, 8 on a segment of 10 observations are larger than the 99th percentile of the data. Following this idea, we implement a super scalable short segment (4S) detection algorithm to cluster the points to form a segment when such a phenomenon occurs.
The advantages of our method are fourfold. First, this nonparametric method requires minimal assumption on the noise distribution. Second, it is super fast as the core algorithm requires only $O(n)$ operations to analyze a sequence of $n$ measurements. In particular, it takes less than 2 seconds for our R codes to analyze 272 sequences with a range of about 34,000 measurements. Third, we establish a non-asymptotic theory to ensure the detection of all signal segments. Last but not least, our method can compute the significance level for each detected segment and offer a convenient approach to statistical inference.

\section{Method}
\subsection{Notations and the main idea}
Let $\{X_j\}_{j=1}^n$ be a sequence of random variables such that
\begin{eqnarray}\label{model}
X_j=\mu_j+\v_j,
\end{eqnarray}
where the \emph{height parameter} vector $\bmu=(\mu_1,...,\mu_n)^{\top}$ is sparse and $\{\v_j\}_{j=1}^n$ are random noises with median 0.
Moreover, we assume that the nonzero entries of $\bmu$ are supported in the union of disjoint intervals $\cI=\bigcup_{k=1}^K I_k$ so that
\[\mu_j=\left\{
          \begin{array}{ll}
            \nu_k\ne0, & \hbox{if }j\in I_k\subset \cI  \hbox{ for some } k; \\
             0, & \hbox{if } j\notin  \cI.
          \end{array}
        \right.\]
Here we assume that $I_k=\{\ell_k,\ell_k+1,...,r_k\}$ and $r_k<\ell_{k+1}-1$ for all $k$. Note that such a representation of $\cI$ is unique and used throughout this paper for $\cI$ or its estimator $\hat\cI$. For convenience and without confusion, we use the interval $[\ell_k,r_k]$ to imply the set of integers $\{\ell_k,\ell_k+1,...,r_k\}$ when referred to a set of locations. We call those intervals \emph{segments}. In particular, a \emph{signal segment} is a segment where the height parameter is a nonzero constant. Let $\bmu_I$ denote a subvector of $\bmu$ restricted to $I\subset\{1,...,n\}$. We denote by $|\cS|$ the cardinality of a set $\cS$. In particular, $|I|=r-\ell+1$ for $I=[\ell,r]$. $\bzero$ and $\bone$ denote vectors, $(0,...,0)^{\top}$ and $(1,...,1)^{\top}$, respectively.

Naturally, a primary goal for model (\ref{model}) is to identify the set of signal segments $\{ I_k\}_{k=1}^K$. Moreover, while rarely done, it is useful to assign a significance level for each of the detected segments. In this paper, we will study both estimation and related inference problems on segment detection. Our strategy is to cluster ``putative locations'' using spatial information. In particular, we consider the set of positions $\cS_c=\{j:|X_j|>c, 1\leq j\leq n\}$, where the observations exceed a threshold $c>0$. Intuitively, for a properly chosen $c$ and some segment $\hat I$, if $|\hat I\cap\cS_c|$ is big enough compared with $|\hat I|$, it is likely that $\bmu_{\hat I}\ne\bzero$.

\begin{figure}[!h]
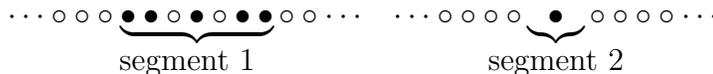

\begin{center}
\[\begin{matrix}
\cdots\circ\circ\circ \underbrace{\bullet\bullet\circ\bullet\circ\bullet\bullet} \circ\circ\cdots&\cdots\circ\circ\circ\circ\underbrace{\bullet}\circ\circ\circ\circ\cdots\\
                       \mbox{segment 1}                                                 &           \mbox{segment 2}
\end{matrix}\]
\end{center}
\caption{Segment 1 and segment 2.}
\label{fig1}
\end{figure}

To illustrate our idea, we consider a game of ball painting. Suppose that we start with $n$ white balls in a row, and paint the $j$th ball with black color if $|X_j|>c$. Let $m$ be the total number of black balls, which is much smaller than $n$. If we observe that there are a few black balls crowded in a short segment, e.g., `segment 1' illustrated in Figure \ref{fig1}, it is plausible that the height parameter is not zero in the segment. Our proposed algorithm can easily identify those segments. Also it may happen that in a neighborhood there is only a single black ball, e.g., `segment 2' in Figure \ref{fig1}. Then, we may not have strong evidence against $\bmu=0$. To put this intuition into a sound theoretical framework, it is imperative to evaluate the significance for each pattern. In fact, given the numbers of white and black balls in a short segment, we may calculate how likely a certain pattern appears in a sequence of length $n$ with $m$ black balls, when white and black balls are actually randomly placed. We will develop a framework of inference based on this idea in Section 2.3.

\subsection{Algorithm}
To estimate $\cI$, we propose a super scalable short segment (4S) detection algorithm which is described as follows.

Step 1: thresholding. Define $\cS_c=\{j:|X_j|>c\}$. That is, we collect the positions where the observations exceed a threshold.

Step 2: completion. Construct the completion set $\bar\cS_{c,d}$ by the criterion that $j\in \bar\cS_c$ if and only if there exist $j_1$, $j_2\in\cS_c$, $j_1\leq j_2\leq j_1+d+1$ such that $j_1\leq j\leq j_2$. That is, we add the whole segment $[j_1,j_2]$ into the completion set if the gap between $j_1$, $j_2\in\cS_c$ is small enough.

Write $\bar\cS_{c,d}=\bigcup_{k=1}^{\tilde K}\tilde I_k$ where $\tilde I_k =[\tilde \ell_k, \tilde r_k]$ with $\tilde r_k<\tilde\ell_{k+1}-1$. Note that this decomposition is unique.

Step 3: clean up.  We delete $\tilde I_k$ from $\bar\cS_{c,d}$ if $|\tilde I_k|=\tilde r_k-\tilde \ell_k+1\leq h$, and obtain our final estimator $\hat\cI_{c,d,h}=\bigcup_{k=1}^{\hat K}\hat I_k$. That is, we ignore the segments that are too short to be considered.

\begin{figure}[!h]
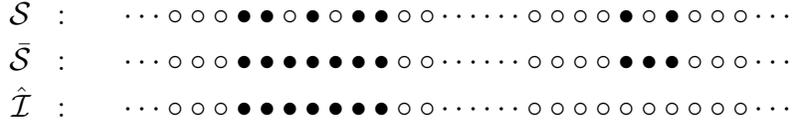

\begin{center}
\begin{eqnarray*}
   \cS&:&     \quad\cdots\circ\circ\circ \bullet\bullet\circ\bullet\circ\bullet\bullet \circ\circ\cdots \cdots\circ\circ\circ\circ\bullet\circ\bullet\circ\circ\circ\cdots\\
   \bar\cS&:& \quad\cdots\circ\circ\circ \bullet\bullet\bullet\bullet\bullet\bullet\bullet \circ\circ\cdots \cdots\circ\circ\circ\circ\bullet\bullet\bullet\circ\circ\circ\cdots\\
   \hat\cI&:& \quad\cdots\circ\circ\circ \bullet\bullet\bullet\bullet\bullet\bullet\bullet \circ\circ\cdots \cdots\circ\circ\circ\circ\circ\circ\circ\circ\circ\circ\cdots\\
\end{eqnarray*}
\end{center}
\caption{An illustration of three steps in the 4S algorithm with $d=h=3$. Top row: the black balls indicate the locations where the observations has absolute value larger than a threshold. Middle row: the small gaps with length $\leq d$ between the black balls are filled in with the black balls. Bottom row: the segment of black balls with length $\leq h$ is deleted.}
\label{fig2}
\end{figure}

The whole procedure depends on three parameters $c$, $d$, and $h$. The choice of $c$ is crucial and depends on applications. $d$ and $h$ are relatively more flexible as we can screen false positives using significance levels defined later. We may ignore the subscripts and simply refer to $\cS$, $\bar\cS$ and $\hat\cI$ when the sets obtained from the three steps above corresponding to $c$, $d$ and $h$ are clear in the context. Figure \ref{fig2} illustrates our procedure, where the location set obtained in each step is indicated by the positions of black balls.

\subsection{Theory: consistency and inference}
\begin{figure}[!h]
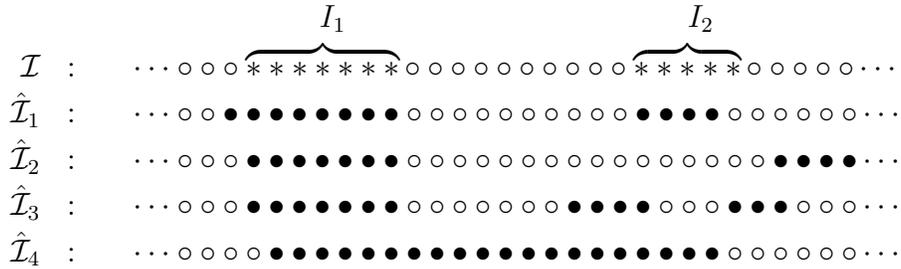

\begin{center}
\begin{eqnarray*}
      & & \qquad\qquad\qquad\quad I_1\quad\qquad\qquad\qquad\qquad\qquad I_2\\
   \cI&:&       \quad\cdots\circ\circ\circ \overbrace{\ast   \ast   \ast   \ast   \ast   \ast   \ast}    \circ\circ\circ\circ\circ\circ\circ\circ\circ\circ \overbrace{\ast\ast\ast\ast\ast}\circ\circ\circ\circ\circ\cdots\\
   \hat\cI_1&:& \quad\cdots\circ\circ\bullet\bullet\bullet\bullet\bullet\bullet\bullet\bullet \circ\circ\circ\circ\circ\circ\circ\circ\circ\circ \bullet\bullet\bullet\bullet    \circ\circ\circ\circ\circ\circ\cdots\\
   \hat\cI_2&:& \quad\cdots\circ\circ\circ \bullet\bullet\bullet\bullet\bullet\bullet\bullet \circ\circ\circ\circ\circ\circ\circ\circ\circ\circ\circ\circ\circ\circ\circ\circ\bullet\bullet\bullet\bullet\cdots\\
   \hat\cI_3&:& \quad\cdots\circ\circ\circ \bullet\bullet\bullet\bullet\bullet\bullet\bullet \circ\circ\circ\circ\circ\circ\circ\bullet\bullet\bullet\bullet\circ\circ\circ\bullet\bullet\bullet\circ\circ\circ\cdots\\
   \hat\cI_4&:& \quad\cdots\circ\circ\circ\circ\bullet\bullet\bullet\bullet\bullet\bullet\bullet\bullet\bullet\bullet\bullet\bullet\bullet
   \bullet\bullet\bullet\bullet\bullet\bullet\bullet\circ\circ\circ\circ\circ\circ\cdots\\
\end{eqnarray*}
\end{center}
\caption{An illustration of relationship between signal segments ($\ast$) and four estimators ($\bullet$). $\cI$ consists of two signal segments $I_1$ and $I_2$. Both $I_1$ and $I_2$ are successfully identified by $\hat\cI_1$. $I_1$ is also identified by $\hat\cI_2$ and $\hat\cI_3$. $\hat\cI_2$ has one true positive (left) and one false positive (right). $\hat\cI_3$ has one true positive (left) and two false positives (middle and right). $\hat\cI_4$ has one false positive.}
\label{fig3}
\end{figure}

Our goal is to identify the set of signal segments $\cI=\bigcup_{k=1}^K I_k$ with a false positive control. Here we say that $I_k\in \cI$ is identified by an estimator $\hat\cI=\bigcup_{k=1}^{\hat K}\hat I_k$ if there is a unique $\hat I_{k'}\in\hat \cI$, such that $\hat I_{k'}\cap I_k\ne\emptyset$, and $\hat I_{k'}\cap I_j=\emptyset$ for all $I_j\in\cI$ and $j\ne k$. Such an $\hat I_{k'}$ is a \emph{true positive}.
We define that $\cI$ is identified by an estimator $\hat\cI$ if every $I_k\in \cI$ is identified by $\hat\cI$. That is, there is a one-to-one correspondence between $\cI$ and a subset of $\hat\cI$, and the $K$ pairs under this correspondence are the only pairs with nonempty interaction among all segments in $\cI$ and $\hat\cI$. See Figure \ref{fig3} for an illustration of the definition.

In our three-step procedure, the first two steps establish an estimator and the last one aims to delete obvious false positives. Our theory proceeds in two main steps. First, we characterize the non-asymptotic probability that the first two steps produce an estimator which successfully identifies $\cI$. In order to identify $I_k\in\cI$, we should ensure two conditions. Condition one is that, after step 1, the black balls are dense enough on $I_k$ so that they do not split into two or more segments in step 2. Condition two is that, in the gap between $I_k$ and $I_{k+1}$, the black balls are sparse enough so that the black balls on $I_k$ and $I_{k+1}$ do not connect to a big segment. Theorem 1 addresses how to bound the probabilities of these two conditions for all $k$.

Second, we develop a framework of inference to control false positives. As a rough control, Theorem 2 gives an upper bound for the expected number of false positives if all segments of length one are deleted in step 3. In general, after step 2, it is not optimal to decide the likelihood of a detected segment being a false positive only by its length. Therefore, for each segment in $\bar \cS$, we check its original color pattern back in step 1 and calculate a $p$-value of this pattern under null hypothesis $\bmu=\bzero$. This assigns a significance level for each detected segment which helps control false positive. It is difficult to find the exact $p$-values. Lemma 2 offers a reasonable approximation.

To facilitate theoretical analysis, we assume that, in this subsection, $\{\v_j\}_{j=1}^n$ are independent and identically distributed (IID) noises with median 0. Moreover, $\v_j$ has a continuous density function $f$ that is symmetric with respect to 0. Under this assumption, the black balls are randomly distributed for arbitrary threshold $c$ when $\bmu=\bzero$.

Now we investigate when a signal segment can be detected by our algorithm. Let $F$ be the cumulative distribution function of the noise density $f$. We use $f_{\alpha}=F^{-1}(\alpha)$ to denote the $\alpha$-percentile. Suppose that there is a segment $I$ such that $|I|=L$, $\bmu_I=\nu\bone$ and $\bmu_{I^c\cap H}=\bzero$, where $H$ is a segment containing $I$ such that $I^c\cap H$ is the union of two segments both of which are of length $D$. Without loss of generality, we assume that $\nu=f_{\alpha}>0$ i.e. $\alpha>0.5$. For a threshold $0<c=f_{\beta}\leq \nu$, let us continue our game of ball painting and focus on this segment and its neighborhood. Recall that we paint the ball at position $j$ with black color if and only if $|X_j|>c=f_{\beta}$. The following two events together can ensure that the segment $I$ is identified by our method.
\begin{eqnarray*}
  \cA &=& \{\text{On }I,\text{ there does not exist a sub-segment of } \min\{d,L\} \text{ consecutive white balls}\}\\
  \cB_D &=& \{\text{On both sides of }I,\text{ within distance } D, \text{ there are } d \text{ consecutive white balls} \}
\end{eqnarray*}
Event $\cA$ ensures that this segment can be detected as a whole segment while event $\cB_D$ controls the total length of the detected segment and makes sure that the detected segments are separated from each other. $\cA$ and $\cB_D$ together guarantee that our algorithm identifies a segment $\hat I$ such that $\hat I\cap I\ne\emptyset$, and $\hat I\cap I'=\emptyset$ for any other signal segment $I'$. The following lemma gives non-asymptotic bounds for $P(\cA)$ and $P(\cB_D)$.

\begin{lemma}\label{lemma1}
Let $H\supset I$ be two segments such that $\bmu_I=\nu\bone$ and $\bmu_{I^c\cap H}=\bzero$, $|I|=L$. $I^c\cap H$ is the union of two segments both of which are of length $D$. Let $\beta'=2\beta-1$.
For $0<c=f_{\beta}\leq \nu$ and $d>0$, after the thresholding and completion steps, we have
                                   \[P(\cA)\geq\left\{
                                      \begin{array}{ll}
                                        1-(L-d+2)2^{-d-1}, & \hbox{if } d<L \\
                                        1-2^{-L}, & \hbox{if } d\geq L
                                      \end{array}
                                    \right.\] and \[P(\cB_D)\geq 1-2(1-\beta'^d)^{\lfloor\frac{D}{d}\rfloor}.\]
\end{lemma}

Let $\nu_{\min}=\min_{k}|\nu_k|$ be the minimal signal strength among all $I_k$'s, $L_{\min}=\min_k |I_k|$, $L_{\max}=\max_k |I_k|$ be the minimal and maximal lengths of signal segments, respectively, and $D_{\min}=\min_k(\ell_{k+1}-r_k-1)$ be the minimal gap between two signal segments. Define $\beta_{\min}=F(\nu_{\min})$ so $\nu_{\min}=f_{\beta_{\min}}$. Let $\beta'_{\min}=2\beta_{\min}-1$. Taking into account all signal segments in $\cI$, the theorem below gives a lower probability bound for identifying $\cI$ after first two steps.

\begin{theorem}\label{thm1}
With $c=\nu_{\min}=f_{\beta_{\min}}$, $d>0$ and $h=0$, $\hat\cI_{c,d,h}$ can identify all signal segments in $\cI$ with probability at least
\begin{eqnarray}\label{prob}
1-K\max\{\frac12(L_{\max}-d+2),1\}2^{-\min\{d,L_{\min}\}}-(K-1)(1-\beta_{\min}'^{d})^{\lfloor\frac{D_{\min}}{d}\rfloor}.
\end{eqnarray}
\end{theorem}

\begin{corollary}\label{cor1}
The probability (\ref{prob}) goes to 1 asymptotically if $\log K+\log L_{\max}\ll \min\{d,L_{\min}\} \to \infty$ and $\log K\ll D_{\min}/d\to\infty$ as $n\to\infty$.
\end{corollary}

Although Theorem 1 gives a theoretical guarantee to recover all signal segments with a large probability, there are some false positives. As an ad-hoc way, we may take $h=2$ or 3 to eliminate some obvious false positives. This clean up step is simple and helpful to delete isolated black balls. The Theorem below gives an upper bound on the number of false positives with a conservative choice $h=1$.

\begin{theorem}\label{thm2}
Assume $\bmu=0$ and $|\cS_c|=m$. Then $E|\hat\cI_{c,d,h}|\leq m (1-\prod_{k=1}^d\frac{n-m+1-k}{n-k})$ with $h=1$.
\end{theorem}

The expected number of false positive segments can be well controlled if both $m/n$ and $d$ are small. Next, we illustrate how to access the significance levels for the detected segments by our method, which is helpful to control false positives. Recall our estimator $\hat\cI=\bigcup_{k=1}^{\hat K}\hat I_k$ where $\hat I_k=[\hat \ell_k,\hat r_k]$. For each $\hat I_k$, let $s_k=|\hat I_k|$, $t_k=|\hat I_k \cap \cS_c|$, and $m=|\cS_c|$. Now consider $n$ balls in a row with $m$ black and $n-m$ white balls. Let $\cA_{n,m,s_k,t_k}$ be an event that there exists a segment of length $s_k$ where at least $t_k$ balls are black, in a sequence of $n$ balls with $m$ blacks ones. The $p$-value of $\hat I_k$ can be defined as the probability of $\cA_{n,m,s_k,t_k}$ if the balls are randomly placed. This $p$-value can effectively control the false positives. However, it is challenging to find the exact formula to calculate the $p$-value. The lemma below gives an upper bound of $P(\cA_{n,m,s_k,t_k})$.

\begin{lemma}\label{lemma2}
$P(\cA_{n,m,s_k,t_k})\leq  \hat P(\cA_{n,m,s_k,t_k})=mP(Y\geq t_k-1)$ where $Y$ follows a hypergeometric distribution with total population size $n-1$, number of success states $m-1$, and number of draws $s_k-1$.
\end{lemma}

This approximated $p$-value is useful to eliminate false positives.

\subsection{Implementation}
Our proposed method is nonparametric and depends on only the rank of absolute measurements $\{|X_j|\}_{j=1}^n$. For a fixed triplet $(c,d,h)$, it typically needs less than $3n$ operations to determine $\hat \cI$ when $|\cS_c|/n$ is small, say, less than 0.1. We need $2n$ operations to compare each measurement with the threshold to determine $\cS_c$. Let $\bw=(w_1,...,w_m)^{\top}$ be a vector of locations in $\cS_c$ in an ascending order. In the completion step, we compare $\Delta\bw=(w_2-w_1,...,w_m-w_{m-1})^{\top}$ to a threshold. In particular, we declare that $w_i$ and $w_{i+1}$ belong to different segments if and only if $w_{i+1}-w_i>d$. Let $i_1$,...,$i_{\tilde K}$ be those indices such as $w_{i_k+1}-w_{i_k}>d$. $\bar \cS$ consists of segments $[w_1,w_{i_1}]$,...,$[w_{i_{\tilde K}},w_m]$. We record the start and end points of each segment only. In the deletion step, we delete $[w_{i_k},w_{i_{k+1}-1}]$ if its length is not greater than $h$. The total operations can be controlled within $2n+10m$.

\begin{figure}[!h]
\begin{center}
\includegraphics[width=5in,height=3in]{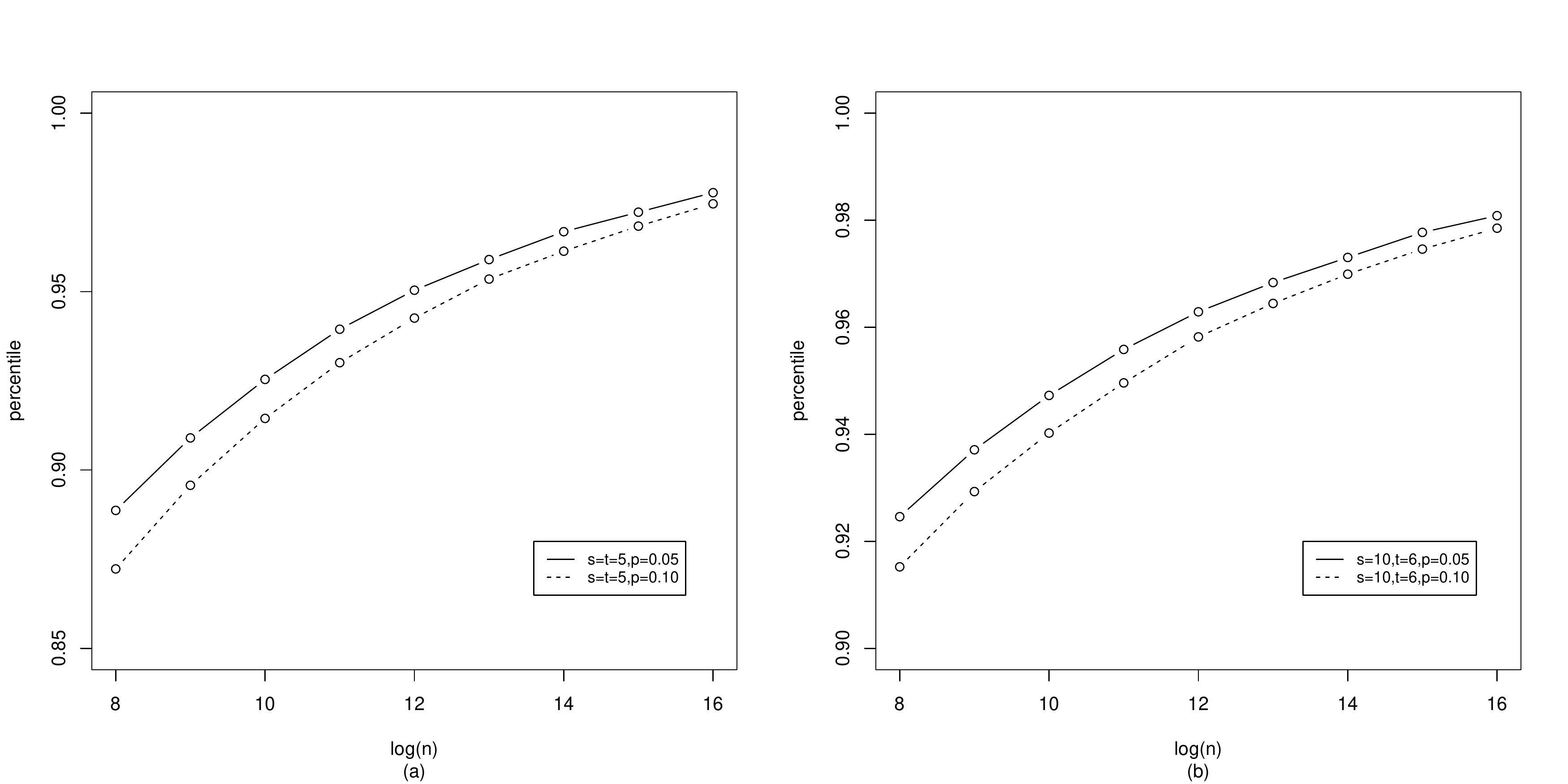}
\end{center}
\caption{Selected percentile versus $\log n$ for (a) $s=t=5$, $p=0.05$ and 0.10; (b) $s=10,$ $t=6$, $p=0.05$ and 0.10.}
\label{fig4}
\end{figure}

The choice of threshold $c$ is crucial to the 4S algorithm, and may need to be determined on a case-by-case basis. Here we offer a general guideline for parameter selection. Recall that for the Gaussian model, the signal strength of a segment with length $L$ and height $\nu=\delta\sigma$ is usually measured by $S=\delta^2L$; see, e.g. Table 1 in \cite{niu2016multiple}. If there are two segments with the same overall signal strength $S$, however, one with large $\delta$ and small $L$ (say, type A), and another one with small $\delta$ and large $L$ (say, type B), then it is usually not equally easy to detect both of them by an algorithm of complex $O(n)$. Indeed, for many segment detection algorithms, it is tricky to balance the powers to detect these two types of segments. Intuitively, the threshold parameter $c$ controls this tradeoff in our methods. A higher threshold may be more powerful in detecting type A segments but less powerful in detecting type B segments; and vice versa. In practice, we may choose the threshold as a certain sample percentile of the absolute values of the observations based on a pre-specified preference. For example, if we know the signal segments have relatively large height parameters but can be as short as 5 data points, then with a fixed $n$, we can find largest $m$ such as $\hat P(\cA_{n,m,5,5})\leq 0.05$ and the percentile is chosen as $\alpha=1-\frac{m}{n}$. That is, we want to guarantee that a segment of 5 consecutive black balls is significant enough to stand out. In another scenario, our preference might be longer segments with possibly lower heights. Then we may choose a threshold to include segments of length 10 with at least 6 black balls. In general, such $m$ (or $\alpha$) can be easily determined given $n$, $s$, $t$ and $p$, by solving $\hat P(\cA_{n,m,s,t})\leq p$. We illustrate in Figure \ref{fig4} the relationship between $\log n$ and selected percentile $\alpha$ for $(s,t)=(5,5)$, $(10,6)$, $p=0.05$ and 0.1. Because our main goal in this paper is to identify short segments, we prefer a large threshold such as 95th sample percentile. An even larger threshold can be used to identify shorter segments, and a smaller threshold can be used for detecting longer segments with lower heights.

\section{Numerical Studies}
\subsection{Simulated data}
We use simulation studies to evaluate the performance of our method in terms of the average number of true positives (TP) and false positives (FP) for identifying signal segments. Recall that in our definition, a detected segment $\hat I\in\hat\cI$ is a true positive, if it interacts with only one signal segment $I\in\cI$, and it is the only one in $\hat\cI$ that interacts $I$.

In Example 1, we show the effectiveness of our inference framework on the false positive control of the 4S algorithm by a null model. As suggested by Figure \ref{fig4}, we choose the 95th percentile of absolute values of the observations as the threshold $c$. We set $d=9$ and $h=3$. We use various $p$-value thresholds for false positive control and compare them with a vanilla version of 4S, which is the one without $p$-value control.
\begin{example}[Null Model]
We generate a sequence based on model (\ref{model}) with $n=10,000$ and $\bmu=\bzero$. We consider three scenarios for the error distributions. In the first two scenarios, we consider $\{\v_i\}_{i=1}^n$ which are IID from $N(0,1)$ and $t_3$, respectively. In the last scenario, we consider $\{\v_i\}_{i=1}^n$ which are marginally $N(0,1)$ and jointly from an autoregressive (AR) model with autocorrelation 0.2.
\end{example}

As $\cI=\emptyset$ in this example, all detected segments are FPs. We report the average FPs for three versions of 4S (Vanilla, $p=0.05$ and $p=0.1$) based on 100 replicates in Table \ref{tab1}. We see that our inference framework can effectively control the number of FPs.

\begin{table}[ht]
\caption {Average number of FPs for the null model} \label{tab1}
\centering
\begin{tabular}{l|ccc}
  \hline
 & 4S (Vanilla) & 4S (p=0.05) & 4S (p=0.1) \\
  \hline
N(0,1) & 102.38 & 0.03 & 0.13 \\
  $t_3$ & 101.68 & 0.12 & 0.26 \\
  AR(1) & 100.39 & 0.10 & 0.33 \\
   \hline
\end{tabular}
\end{table}

In Example 2, we compare 4S with three algorithms CBS \citep{olshen2004circular}, LRS \citep{jeng2010optimal} and WBS \citep{fryzlewicz2014wild}. The CBS and WBS methods, implemented by \texttt{R} packages \texttt{DNAcopy} and \texttt{wbs} respectively, give a segmentation of the sequence which consists of a set of all segments rather than only the signal segments. In order to include their results for comparison, we ignore the long segments (with length greater than 100) detected by CBS or WBS, which decreases their false positives. For LRS, we set the maximum length of signal segments as 50.

\begin{example}
We generate a sequence based on model (\ref{model}) with $n=10,000$. There are 5 signal segments with lengths 8, 16, 24, 32 and 40 respectively. We use the same error distributions as in Example 1. We consider two levels of height parameter for different signal strengths. In particular, we set height $\nu$ as the 99-, and 97-th percentiles of the marginal error distribution in two scenarios, labeled by S1 and S2. For the standard normal error, the height values are 2.326 and 1.881, respectively.
\end{example}

The threshold $c$ we used for 4S is the 95-th sample percentile of absolute values, that is around the 97.5-th percentile of the error distribution, e.g., around 1.96 for the Gaussian case. Therefore, the true height is greater than $c$ in S1, but lower in S2. Average numbers of TPs and FPs are reported in Tables \ref{tab2} and \ref{tab3}.

\begin{table}[ht]
\caption {Average number of TPs} \label{tab2}
\centering
\begin{tabular}{l|cccccc}
  \hline
S1 & 4S (p=0.05) & 4S (p=0.1) & 4S (p=0.5) & CBS & LRS & WBS \\
  \hline
N(0,1) & 4.41 & 4.58 & 4.73 & 4.89 & 4.41 & 4.53 \\
$t_3$ & 4.95 & 4.98 & 4.99 & 2.16 & 4.41 & 4.52 \\
AR(1) & 4.40 & 4.53 & 4.65 & 4.68 & 4.41 & 4.53 \\
  \hline
S2 & 4S (p=0.05) & 4S (p=0.1) & 4S (p=0.5) & CBS & LRS & WBS \\
  \hline
N(0,1) & 3.77 & 3.94 & 4.20 & 4.59 & 3.76 & 3.94 \\
$t_3$ & 3.34 & 3.47 & 3.73 & 0.51 & 3.75 & 3.93 \\
AR(1) & 3.75 & 3.94 & 4.09 & 4.43 & 3.76 & 3.95 \\
  \hline
\end{tabular}
\end{table}

\begin{table}[ht]
\caption {Average number of FPs} \label{tab3}
\centering
\begin{tabular}{l|cccccc}
  \hline
S1 & 4S (p=0.05) & 4S (p=0.1) & 4S (p=0.5) & CBS & LRS & WBS \\
  \hline
N(0,1) & 0.02 & 0.05 & 0.29 & 0.10 & 0.05 & 0.14 \\
 $t_3$ & 0.04 & 0.09 & 0.36 & 0.13 & 0.45 & 0.36 \\
 AR(1) & 0.05 & 0.14 & 0.44 & 1.99 & 0.05 & 0.14 \\
    \hline
S2 & 4S (p=0.05) & 4S (p=0.1) & 4S (p=0.5) & CBS & LRS & WBS \\
  \hline
N(0,1) & 0.02 & 0.08 & 0.40 & 0.16 & 0.09 & 0.22 \\
 $t_3$ & 0.10 & 0.18 & 0.63 & 0.07 & 0.49 & 0.44 \\
 AR(1) & 0.09 & 0.22 & 0.58 & 1.84 & 0.09 & 0.23 \\
    \hline
\end{tabular}
\end{table}

Overall, CBS performs the best for the IID Gaussian case, but suffers from a low power in the heavy-tail case, and high FPs in the correlated case. LRS and WBS perform reasonable well with slightly high FPs in the heavy-tail case. 4S methods are more robust against the error type. When the noise is Gaussian and the signal strength is weak, it is slightly less powerful than the methods based on the Gaussian assumption. In terms of computation time (Table \ref{tab3.5}), 4S is about 100 times faster than all other methods.

\begin{table}[ht]
\caption {Computation time (in second) to complete 300 sequences in S1 for each method.} \label{tab3.5}
\centering
\begin{tabular}{|c|cccc|}
  \hline
  Method & 4S & CBS & LRS & WBS \\
  Time & 0.83 &115.52 &108.72 & 86.17 \\
  \hline
\end{tabular}
\end{table}

\subsection{Real data example}

We applied the 4S method to the 272 individuals from HapMap project. In particular, we tried 4S (with $p$-value thresholds 0.05 and 0.5) to the LRR sequence of chromosome 1, which consists of 33991 measurements for each subject. We compared 4S with CBS, which has been a benchmark method in CNV detection. Note that CBS produces a segmentation of the sequence rather than the CNV segments directly. Therefore, we focused on only the short (less than 100 data points) segments detected by CBS because the long segments had means close to zero and are not likely to be CNVs. We found that most of these short segments are separated. But a very small portion of them are connected as CBS sometimes tends to over segment the sequence. Therefore, we merged two short segments detected by CBS if they are next to each other.

The 4S algorithm is extremely fast. It took less than 2 seconds (on a desktop with CPU 3.6 GHz Intel Core i7 and 16GB memory) to complete 272 sequences with $p$-values calculated for all detected segments. CBS algorithm is reasonably fast, but much slower than our algorithm. In Table \ref{tab4} we list the total number of detected CNVs, average length of CNVs and computation time for all methods.

\begin{table}[ht]
\caption {Real data results: total number of detected CNVs, average length of CNVs, and computation time for all methods.} \label{tab4}
\centering
\begin{tabular}{|l|ccc|}
  \hline
                 & 4S (p=0.05) & 4S (p=0.5) & CBS \\
                  \hline
  Number of CNVs & 2832 & 3141 & 2962 \\
  Average length & 28.60 & 27.70 & 23.15 \\
  Computation time & 1.86 & 1.93 & 230.18 \\
  \hline
\end{tabular}
\end{table}

Overall, the segment detection results were very similar. We further compared the segments detected by two algorithms, i.e., 4S with threshold $p=0.05$ and CBS. We found that 2753 segments are in common. Here by a common segment we mean a pair of segments, one detected from each algorithm, such that they overlap to each other but do not overlap with other detected segments. Among these common segments, we calculated a similarity measure, called affinity in \cite{arias2005near}, defined as follows.
\begin{align}\label{f3.1}
\rho(I,I')=\frac{|I\cap I'|}{\sqrt{|I|\cdot|I'|}}.
\end{align}
$\rho(I,I')=1$ if two segments $I$ and $I'$ are the same and $\rho(I,I')=0$ if they do not overlap. We found that the average value of this similarity measure is 0.9290 among 2753 pairs.

\begin{figure}[!h]
\begin{center}
\includegraphics[width=2.6in,height=2.6in]{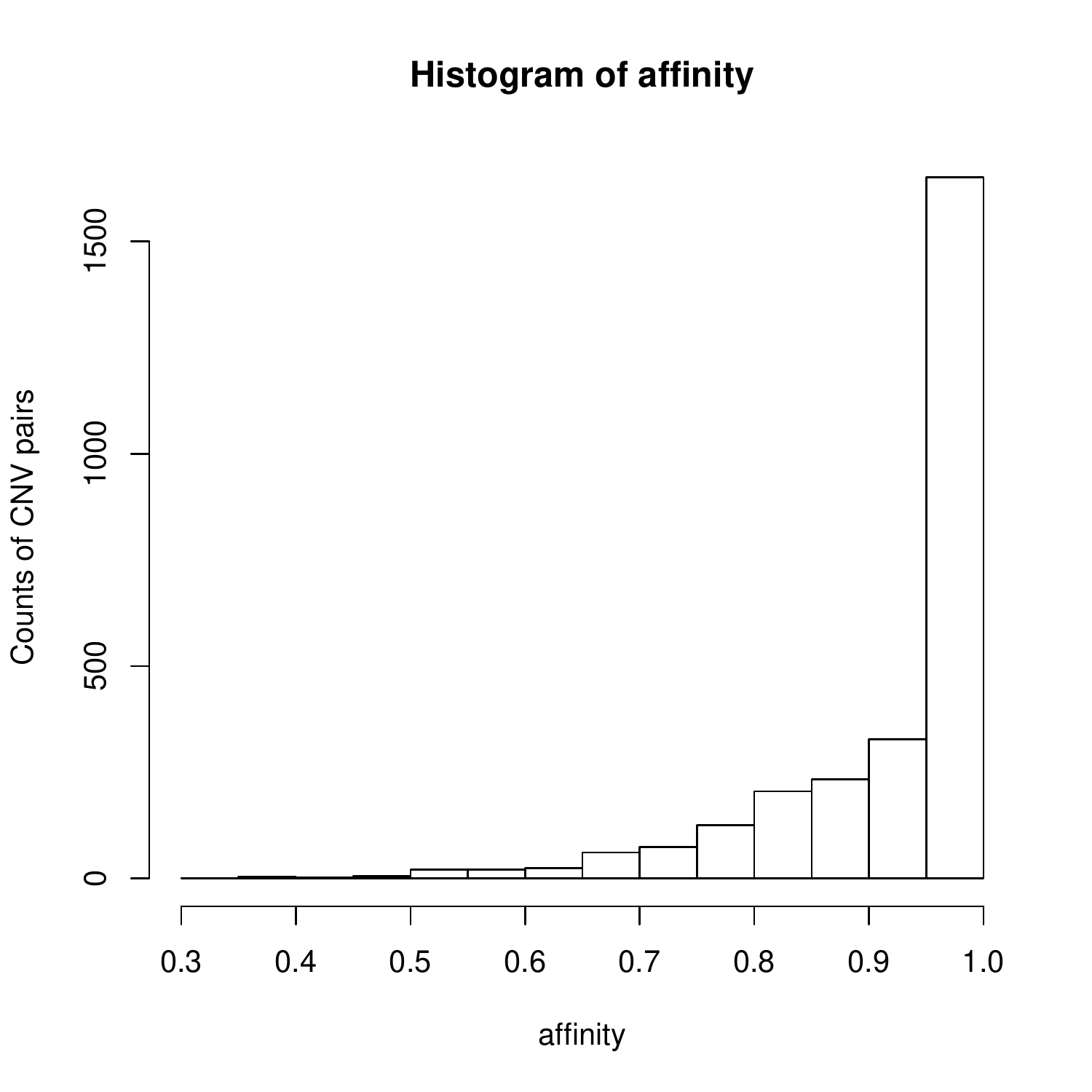}                    
\end{center}
\caption{Histogram of affinity, a similarity measure defined in (\ref{f3.1}), among 2753 pairs of commonly detected CNVs by 4S and CBS. Affinity equals to 1 if two detected CNVs are identical, and equals to 0 if two detected CNVs do not overlap. }
\label{fig5}
\end{figure}

\begin{figure}[!h]
\begin{center}
\includegraphics[width=5in,height=5in]{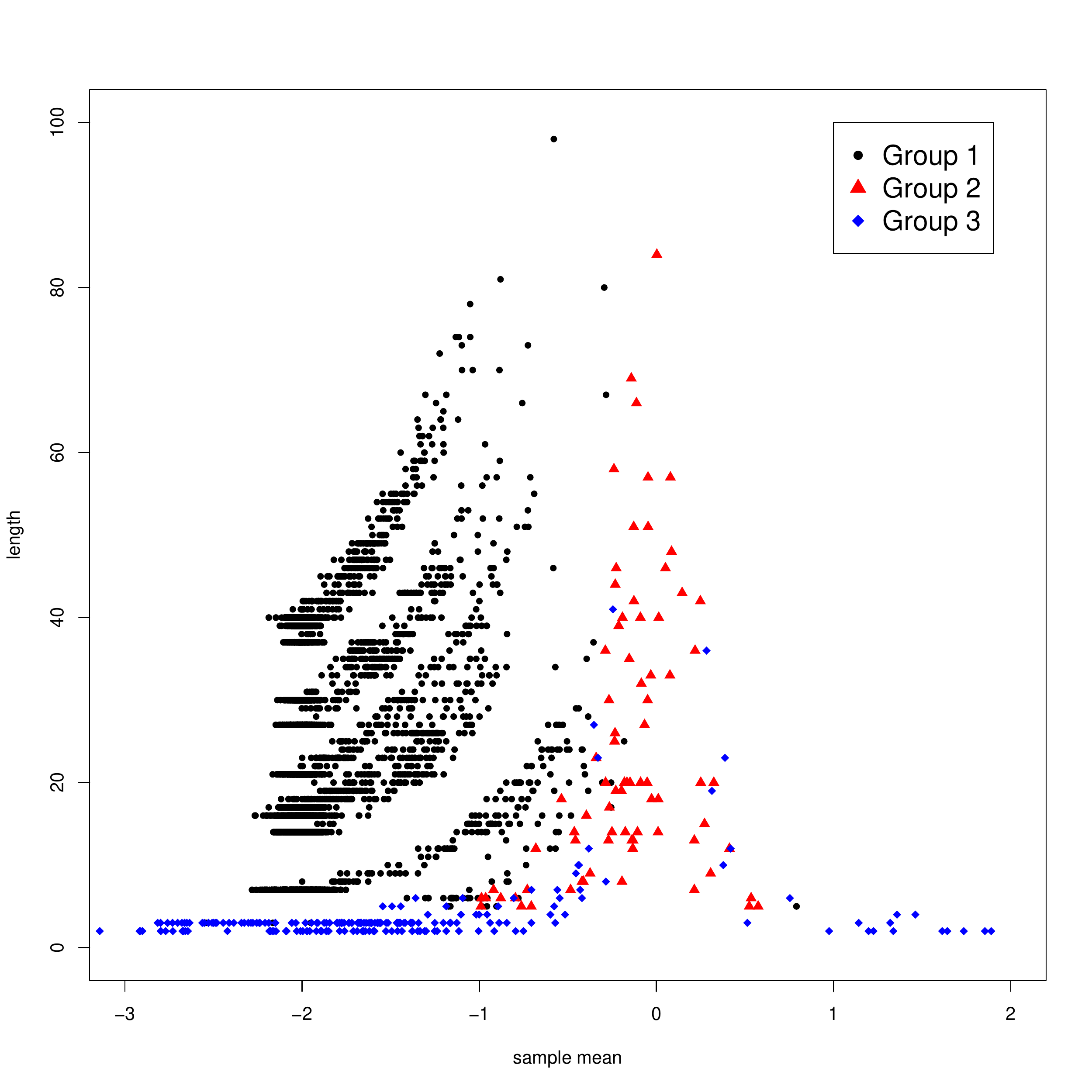}                    
\end{center}
\caption{Scatter plots of sample means versus lengths for detected segments in three groups. Group 1: segments detected by both methods; Group 2: segments detected by only 4S method with $p=0.05$; Group 3: segments detected by only CBS.}
\label{fig6}
\end{figure}

Figure \ref{fig5} presents the histogram of affinity among 2753 pairs of commonly detected segments by 4S and CBS. We can see that 87.76\% of those pairs have affinity values larger than 0.8. We further divided the detected segments into three groups: those detected by both methods (group 1); those detected by only 4S method with $p=0.05$ (group 2); those detected by only CBS (group 3). For each detected segment, we calculated its length and the sample mean of the measurements on the segment. Figure \ref{fig6} displays the scatter plots of sample means versus lengths for all the segments in three groups. Most segments in group 1 carries relatively strong signals. So it is not surprised that they were detected by both algorithms. The groups 2 and 3 have much smaller sizes than group 1. In particular, we found that most segments in group 3 (i.e. those detected by only CBS) are very short, consisting of only 2 or 3 data points. Those segments are not significant in our inference framework unless we set a very high threshold $c$ in step 1. Some segments in group 2 (i.e. those detected by only the 4S method) have relatively small sample mean values, which explains why they were not detected by CBS. Some of these segments might be true positives with the sample mean affected by outliers. Overall, the 4S and CBS methods gave similar results. The segments detected by only one method may be prone to false positives, or true positives have weak signal strengths.

\section{Discussion}
We proposed a scalable nonparametric algorithm for segment detection, and applied it to real data for CNV detection. Two main advantages of the 4S algorithm are its computational efficiency and independence of the normal error assumption. We introduced an inference framework to assign significance levels to all detected segments. Our numerical studies demonstrated that our algorithm was much faster than CBS and performed similarly to CBS under the normality assumption and better when the normality assumption was violated. Although our inference framework depended on the assumption of IID noise, our numerical experiments suggested that our algorithm worked well under weakly correlated noises. Hence, the proposed method is faster and more robust against non-normal noises than CBS. Overall, the 4S algorithm is a safe and fast alternative to CBS, which has been a benchmark method in CNV studies.

In the literature, there are two popular classes of change-point models used to study CNV related problems. The first one assumes only a piecewise constant median/mean structure. The second one assumes, in addition, a baseline, which reflects the background information or normal status of the data. Quite often, it assumes that the abnormal part, called signal segments in our paper, are sparse. For the first approach, the goal is to identify the change points. In contrast, the second approach emphasizes more on segment detection rather than change-point detection. The difference is subtle for estimation but might become remarkable for inference. For example, it is technically difficult to define `true positive' in the context of change-point detection \citep{hao2013multiple}. But it is easier to define related concepts for segment detection as we did in this paper. Roughly speaking, the first approach is more general, and the second one is more specific and suitable to model certain CNV data, e.g., SNP array data. In particular, the 4S algorithm aims to solve change-point models in the second class. It can be applied to any data sequence when there is a baseline. When the baseline mean/median is unknown, we suggest that the data should be centered first by the estimated mean/median. Our method can not be applied to data when a baseline does not exist. Besides change-point models, there are other approaches to study CNV such as hidden Markov model \citep{wang2007penncnv}. Due to the space limit, we restricted our comparison to the methods based on change-point models and implemented by R packages.

Most segment detection algorithms involve one or more tuning parameters, whose values are critical to the results. In the study of segment detection, there are two trade-offs that researchers should consider in choosing algorithms as well as their parameters. The first one is the usual type I/type II errors trade-off, which might be tricky sometimes but well-known. The second one is more delicate and quite unique. For a signal segment, both its height and length determine the signal strength. Therefore, segments with weak but detectable signals can be roughly divided into two categories, the ones with small length (say, type A) and the ones with small height (say, type B).
Typically a method may detect type A segments more powerfully, but type B segments less powerfully, than the other method. For the proposed 4S algorithm, a choice of a larger threshold parameter in step 1 makes the algorithm more powerful in detecting short and high signal segments (type A), and vice versa. The 4S algorithm can be easily tuned to maximize the power in detecting of a certain type of the signal segments. We may also try different thresholding levels in data analysis in order to detect different types of segments. In general, the choice of the parameters depends on the research goals and balance of two trade-offs mentioned above.

There are various platforms and technologies which produce data for CNV detection. Besides the SNP array data studied in this work, read depth data from next generation sequencing (NGS) technologies are often used in CNV studies. As one referee pointed out, the speed of 4S algorithm would be an advantage when applied to read depth data from whole genome sequencing. This is a wonderful research direction that we will investigate next.

An R package \texttt{SSSS} implementing our proposed method can be download via\\
\texttt{https://publichealth.yale.edu/c2s2/software}.

\section*{Acknowledgement} The authors are partially supported by National Science Foundation, National Institutes of Health, Simons Foundation and University of Arizona faculty seed grant.

\section{Appendix}

\noindent\textbf{Proof of Lemma 1.}\\
Let $I$ be the interval of integers $[\ell,r]$ with $L=r+1-\ell$. For each $X_i$, $i\in I$, the probability that the ball at $i$ is white is $\pi=P(|X_i|\leq c)\leq P(X_i\leq c)\leq \frac12$ as $\nu\geq c$ and $f$ is symmetric. It is trivial to bound $P(\cA)$ for the case $d\geq L$ as $P(\cA^c)=\pi^{L}\leq 2^{-L}$.  Now let us consider the case $d<L$. Let $\cE_i$, $i\in I$ be the event that the first segment of $d$ consecutive white balls starts from position $i$. Then
\[P(\cE_i)\left\{
            \begin{array}{ll}
              =\pi^d, & \hbox{if } i=\ell; \\
              \leq (1-\pi)\pi^d, & \hbox{if } \ell<i\leq r+1-d \\
              =0, & \hbox{if } i\geq r+2-d.
            \end{array}
          \right.\]
Therefore, $P(\cA^c)= \sum_{i\in I} P(\cE_i) \leq \pi^d+ (L-d)(1-\pi)\pi^d \leq (1+\frac12(L-d))(\frac12)^d$ and $P(\cA)\geq 1- (L-d+2)2^{-d-1}$. Note that $\pi$ is a constant depending on $f$, $\nu$ and $c$, so a sharper bound than $\frac12$ for $\pi$ may be used to bound $P(\cA)$ if more information is available.

Let us consider the segment $[r+1,r+D]$ on the right hand side of $I$. $\cB_{D}^c$ implies that there is at least one black ball in each of the segments $[r+1,r+d]$, $[r+d+1,r+2d]$, etc. Note that $X_i\sim F$ on these segments so the probability of white ball at $i$ is $P(|X_i|\leq c=f_{\beta})=\beta'=2\beta-1$. Consider all $\lfloor \frac{D}{d}\rfloor$ segments of length $d$ on the right side of $I$. The probability that all these segments contain at least one black ball is $(1-\beta'^d)^{\lfloor\frac{D}{d}\rfloor}$. Therefore, $P(\cB_D)\geq 1-2(1-\beta'^d)^{\lfloor\frac{D}{d}\rfloor}$. $\square$

\noindent\textbf{Proof of Theorem 1.}\\
For $\cI=\cup_{k=1}^K I_k$, let $L_k=|I_k|$ and $D_k$ be the gap between $I_k$ and $I_{k+1}$. Define
\begin{eqnarray*}
  \cA_k &=& \{\text{on }I_k,\text{ there does not exist a sub-segment of } \min\{d,L_k\} \text{ consecutive white balls}\}\\
  \cB_k &=& \{\text{there are } d \text{ consecutive white balls on } D_k \}.
\end{eqnarray*}
Note that all segments in $\cI$ are identified under event $\left(\bigcap_{k=1}^K\cA_k\right)\bigcap\left(\bigcap_{k=1}^{K-1}\cB_k\right)$. By Lemma \ref{lemma1}, $P(\cA_k^c)\leq 2^{-L_k}$ or $\frac12(L_k-d+2)2^{-d}$, which can be bounded by $\max\{\frac12(L_{\max}-d+2),1\}2^{-\min\{d,L_{\min}\}}$. Moreover, $P(\cB_k^c)\leq (1-\beta_{\min}'^d)^{\lfloor\frac{D}{d}\rfloor}\leq (1-\beta_{\min}'^d)^{\lfloor\frac{D_{\min}}{d}\rfloor}$. The conclusion follows Bonferroni inequality. $\square$

\noindent\textbf{Proof of Theorem 2.}\\
Let $\cS_c=\{j_1,...,j_m\}$ be the locations of black balls after step 1. Note that $j_i$ and $j_{i+1}$ will be connected in step 2 if and only if $j_{i+1}-j_i\leq d$. We aims to count the number of segments with at least 2 consecutive black balls after step 2, as all isolated black balls will be eliminated in step 3. Such a segment starts at $j_i$ only if $j_{i+1}-j_i\leq d$. So the total number of such segments is at most $|\{i: j_{i+1}-j_i\leq d\}|$. Let $Z_i$ follow Bernoulli distribution with $Z_i=1$ if and only if $j_{i+1}-j_i\leq d$ for $i=1,...,m$. When $\bmu=\bzero$, all black balls are randomly distributed. $P(Z_i=0)$, i.e., the probability that all balls are white in next $d$ positions following $j_i$ is ${n-1-d\choose m-1}/{n-1\choose m-1}=\prod_{k=1}^d\frac{n-m+1-k}{n-k}$. Therefore, $E|\hat\cI_{c,d,h}|\leq \sum_{i=1}^m Z_i\leq m (1-\prod_{k=1}^d\frac{n-m+1-k}{n-k})$. $\square$

\noindent\textbf{Proof of Lemma 2.}\\
We drop the subscript $k$ in $P(\cA_{n,m,s_k,t_k})$ as it is irrelevant in our derivation below. Under the assumption that $m$ black balls are randomly assigned in $n$ position, at a position of black ball, we calculate the probability that there are at least $t-1$ black balls in next $s-1$ positions. Let $Y$ be the count of black balls in those $s-1$ positions. $Y$ follows a hypergeometric distribution with total population size $n-1$, number of success states $m-1$, and number of draws $s-1$. Therefore, $P(\cA_{n,m,s_k,t_k})\leq mP(Y\geq t-1)$ as there are $m$ black balls. $\square$

\bibliographystyle{biometrika}
\bibliography{ssss}

\end{document}